\documentstyle[emulateapj,epsfig]{article}

\newcommand{\rsun}{{R}_{\odot}}
\newcommand{\msun}{{M}_{\odot}}

\newcommand{\mdot}{\dot{M}}

\newcommand{\msyr}{\msun \ {\rm yr^{-1}}}

\begin{document}

\title{Discovery of a Magnetic White Dwarf in the Symbiotic Binary Z Andromedae}

\author{J. L. Sokoloski and Lars Bildsten}
\affil{Department of Physics and Department of Astronomy, University of
California, Berkeley, Berkeley, CA 94720-3411; jeno@song.berkeley.edu,
bildsten@fire.berkeley.edu} 
\authoraddr{601 Campbell Hall, U.C. Berkeley, Berkeley, CA 94720-3411;
jeno@song.berkeley.edu} 
\begin{abstract}

\end{abstract}

 We report the first result from our survey of rapid variability in
symbiotic binaries: the discovery of a persistent oscillation at
$P=1682.6\pm0.6$ s in the optical emission from the prototype
symbiotic, Z Andromedae.  The oscillation was detected on all 8
occasions on which the source was observed over a timespan of nearly a
year, making it the first such persistent periodic pulse found in a
symbiotic binary. The amplitude was typically $2 - 5$ mmag, and it was
correlated with the optical brightness during a relatively small
outburst of the system. The most natural explanation is that the
oscillation arises from the rotation of an accreting magnetic ($B_{\rm S} \gtrsim
10^5$G) white dwarf.  This discovery constrains the outburst
mechanisms, since the oscillation emission region near the surface of
the white dwarf was visible during the outburst.


\section{Introduction}

When the term ``symbiotic star'' was coined in the early 1940s for the
newly discovered peculiar variable stars with combination optical
spectra (see \cite{ken86}), Z Andromedae was one of the prototypes.
Today, it remains one of the most frequently observed symbiotic systems
(SS).  The observations have revealed a complex system that is still not
fully understood (\cite{mk96}).  Most evidence indicates that the hot
star in Z Andromedae is a white dwarf (WD), and the work we present here
supports that conclusion. This evidence includes effective temperature
estimates of the hot component of approximately $10^5 {\rm  K}$ (\cite{fc88};
\cite{mur91}), an inferred hot component radius of approximately $0.07
\rsun$ (\cite{fc88}; \cite{mur91}), and a large radio nebula
(\cite{stb}), which is not expected if mass transfer occurs via Roche
lobe overflow onto a main sequence star.  The binary has an orbital
period of 759 days (\cite{fl94}; \cite{mk96}), and Schmid and Schild
(1997) have used Raman line polarimetry to determine an orbital
inclination of $i \approx 47 \pm 12^\circ$ and infer a mass for the hot
component of $0.65 \pm 0.28 \msun$ (assuming a total system mass of
between 1.3 and 2.3 $\msun$).

According to current theories of binary evolution (\cite{yun95}), most
white dwarfs found in symbiotics should have evolved from stars with
main sequence masses greater than about $1.5 \msun$.  Highly magnetic
WDs ($B_{\rm S} \gtrsim 10^6$ G, where $B_{\rm S}$ is the field at the stellar surface)
appear to be preferentially formed by stars with main sequence masses $M
\approx 2-4 \msun$ (\cite{ang81}; \cite{sio88}), and so it is possible
that the fraction of WDs that are magnetic is higher in symbiotics than
in the field, where it is about $3-5$\% (\cite{cha92}).  Given that
there are at least 150 known SS, and that most of these contain white
dwarfs, we expect some SS to contain white dwarfs that are magnetized at
the level seen in DQ Herculis and AM Herculis cataclysmic variables
($B_{\rm S} \gtrsim 10^5$ G). 
Miko{\l}ajewski et al. (1990ab) and also Miko{\l}ajewski \&
Miko{\l}ajewska (1988) have invoked the presence of a magnetic white
dwarf to explain the jets, flickering with possible QPO's, and large changes
in the hot component luminosity in the symbiotic star CH Cygni, and this
idea was later also adopted in the case of another symbiotic, MWC 560 (\cite{tom92};
\cite{mic93}).  However, stable and repeatable oscillations like those
detected in magnetic cataclysmic variables have until now not been seen
in a symbiotic, and the prevalence of magnetic white dwarfs in SS is an
important unknown.

We are undertaking a long-term observational program to study the minute
time scale photometric behavior of symbiotic binaries, expanding upon the
work of Dobrzycka, Kenyon, \& Milone  (1996) and others.  In \S \ref{sec:obs},
we present the first result from our survey, the discovery of a
28-minute oscillation in the B band emission from Z Andromedae.  This
was the only strong oscillation found in the preliminary analysis of 20
objects.  Results from the complete survey will be presented in a future
paper. In \S \ref{sec:magacc}, we interpret this oscillation as due to
accretion onto a magnetic, rotating white dwarf.  The fact that the
oscillation was detected during a recent outburst, as well as once the
source had returned to quiescence, has implications for outburst models,
as we outline in our conclusions (\S \ref{sec:obmodels}).

\section{Observations and Results } \label{sec:obs}

We observed Z Andromedae on seven occasions separated by two to four
weeks each from 1997 July to 1997 December, and then once again in
1998 June, with the 1-meter Nickel telescope at UCO/Lick Observatory.
The observations ranged in length from approximately 4 hours on a single
night in 1997 July, to approximately 7 hours per night three nights in a row
in 1997 November, for a total of 76 hours of observing on 13 nights
spanning 1 year (see Table 1).  The $2048\times 2048$ pixel,
$6.3\times 6.3$ arcmin, unthinned LORAL CCD currently in Lick's dewar
\#2, and a Johnson B filter were always used.  The first observation
fortuitously occurred about 1 month after the peak of an optical
outburst ($\Delta V \approx 1$), and the subsequent observations in
1997 took place as the optical flux declined.  At the time of the 1998
June observation, the optical flux had returned to its pre-outburst
quiescent value.  A 2.2 year V band light curve is shown in Figure 1,
with the times of our observations marked.

\epsfig{file=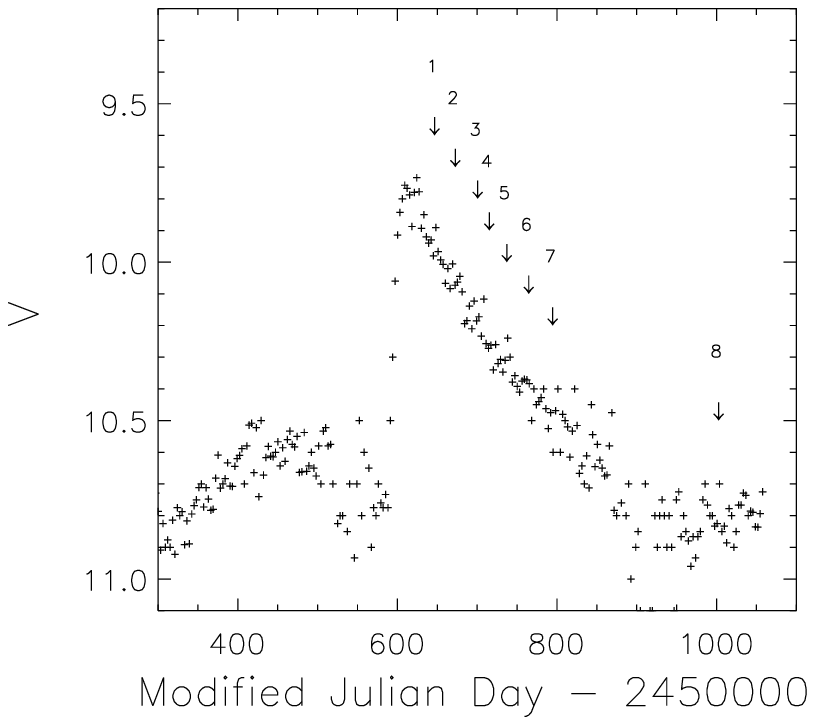}
\figcaption[f1.eps]{Long term V band light curve of Z Andromedae, from the American
Association of Variable Star Observers (\protect\cite{mat98}).  The times
of our 8 observations are marked with arrows; see
Table 1 for the corresponding dates. \label{fig1}}

\vspace{4mm}
At the time of the first observation, on 1997 July 8, the binary was
oriented with the WD in front of the red giant, from the observer's perspective
(i.e. the orbital phase of the binary was 0.5, where phase 0.0
corresponds to photometric minimum in quiescence and spectroscopic
conjunction).  By 1997 December 2, the binary had moved through almost
one quarter of its orbit to phase 0.7, where both stars are roughly
equidistant from the observer, and in 1998 June, the WD was behind the
red giant.

Data reduction was performed using IDL software based on standard IRAF
routines.  Source counts for each image were extracted from a circular
aperture with a radius of 8 to 14 arcsec, and the background was
estimated from a surrounding annulus.  For each light curve, the
extraction region was chosen to be much larger than the seeing so that
any variability due to source counts falling outside the extraction
region (as the seeing or guiding quality changed) was small compared to
systematic errors.  Z Andromedae is bright enough that even with large
extraction regions, the Poisson errors from sky background are usually
not significant.  Several representative light curves from our
observations are shown, in chronological order, in Figure 2.  There is
one other bright star in the field of Z Andromedae (at J2000 coordinates
23 33 24, +48 45 38), but it is also variable, so it was not used as a
comparison star for differential photometry (except for one night in
October when its amplitude of variability was low, and one night in
November when thick clouds were present).  Therefore, although every
attempt was made to perform observations on clear nights, some
observations were affected by high clouds.  Data points affected by
radiation events (``cosmic rays'') were removed when they could be
identified, and the light curves were corrected for atmospheric
extinction.  In addition, we divided most of the light curves by a 3rd
order polynomial in order to remove residual atmospheric effects
($\lesssim$ 1\% effect).  Note that this may have also removed any
intrinsic variability on a time scale comparable to the length of the
observation.  This polynomial fitting was not performed for the Aug. 2
and Aug. 30 data because of the presence of flare-like variability in
the light curves (see Figure 2).

\epsfig{file=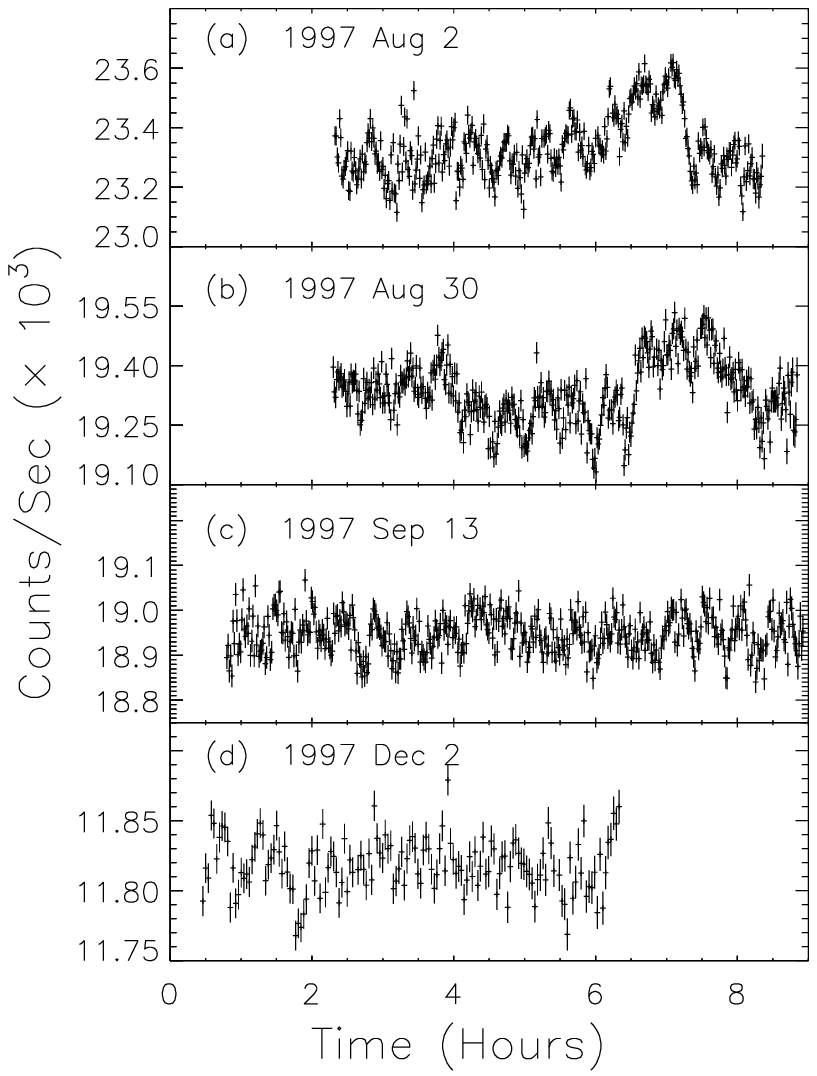}
\figcaption[f2.eps]{Four B Band light curves of Z Andromedae.  The
light curves have been corrected for atmospheric extinction, and
normalized to a typical value of the count rate. The 28-minute
oscillation is clearly visible in at least 3 of the 4 light
curves.  The additional variability on August 2 and August 30 could be
intrinsic to Z Andromedae, but the lack of a constant comparison star
prohibits us from ruling out atmospheric origins.  \label{fig2}}

\vspace{4mm}
Power spectra corresponding to the light curves in Figure 2 are shown in
Figure 3.  The most striking, persistent feature is the peak at 0.6 mHz,
corresponding to a period of 28 minutes.  A smaller, but still
significant peak is also present at twice this frequency in the power
spectra of the July 8 (not shown), August 2, August 30, and September 13
light curves.  The 28-minute oscillation was significantly detected in
all 8 observations.  The power spectrum of the other bright star in the
field does not show the feature at 0.6 mHz, confirming that the
oscillation detected in Z Andromedae was neither instrumental nor
atmospheric in origin.

\epsfig{file=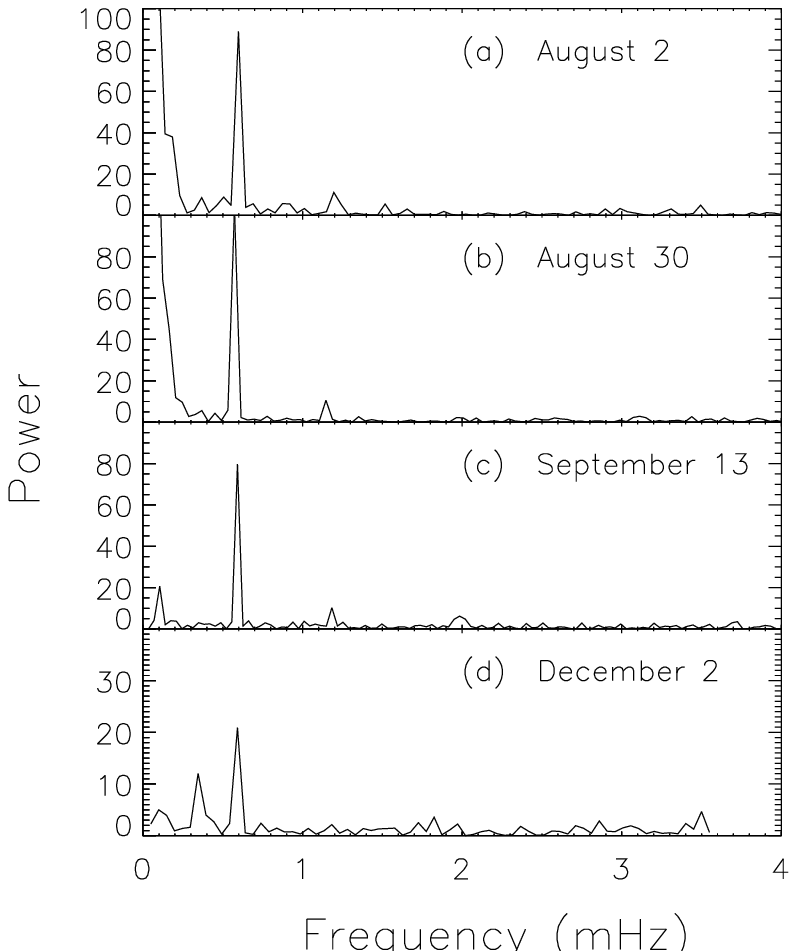}
\figcaption[f3.eps]{Power spectra for the light curves in Figure
2. The power is plotted in units of mean high frequency power (which is
``white''), and the feature at 0.6 mHz corresponds to an oscillation
period of 28 minutes.  A smaller feature at 1.2 mHz is also seen (and
significantly detected) in the top three panels.  The features at
frequencies lower than 0.6 mHz are not repeated from one observation to
another, and could be due to atmospheric changes or source
variability.\label{fig3}}

\subsection{Timing Analysis}

Although the feature at 0.6 mHz is detected in the power spectra of Z
Andromedae, the precise value of the oscillation period and its
uncertainty is best determined in the time domain.  Visual inspection of
the light curves suggests that the signal is not a simple sinusoid, and
the detection of a harmonic in several of the power spectra confirms
this impression.  By using time domain epoch folding techniques, we need
not make any assumptions about the shape of the pulse profile, and all
the signal ``power'' will be located at a single period.

To perform this analysis, we used the phase dispersion minimization
(PDM) technique originally described by Stellingwerf (1978).  The PDM
technique consists of folding a light curve at a range of periods and
computing the mean pulse profile, and the scatter of the data points
about this profile, for each period.  Each data point is assigned a
phase $\phi=t \bmod P$, where $t$ is the time of the measurement from
some initial time and $P$ is the fold period, and binned accordingly.
In our case the number of phase bins ranged from 10 to 20, and was
chosen to be as large as possible while still ensuring that each phase
bin contained enough data for our statistics to be valid (at least 10
data points).  With a large number of points in each bin, the standard
PDM statistic is just $\chi^2$,
\begin{equation}
\chi^2 = \sum_{i=1}^{n_b} \sum_{j=1}^{n_i} \frac{(x_{ij}-m_i)^2}{\sigma_{ij}^2},
\end{equation}
where $n_b$ is the number of phase bins, $n_i$ is the number of points
in bin $i$, $x_{ij}$ is the $j$th point in the $i$th phase bin,
$m_i=n_i^{-1} \sum x_{ij}$ is the mean of all points in bin $i$, and
$\sigma_{ij}$ is the uncertainty on $x_{ij}$.  In our treatment the
$\sigma_{ij}$ were usually dominated by the Poisson errors on the
source counts.  The mean pulse profile will be rather flat for fold periods
far from the true period, and the points in each phase bin will have a
large variance, causing $\chi^2$ to be large.  For a fold period close
to the true period, the mean profile will approach the true pulse
profile, the variance of the points in each bin will be small, and
$\chi^2$ will decrease.  The best estimate of the true period is found
by minimizing $\chi^2$.

For a data stream with Gaussian noise and a superimposed
oscillation, $\chi^2_{min}$, the minimum value of $\chi^2$, should be
approximately equal to the number of degrees of freedom (in
this case $N-n_b$, where $N$ is the total number of points).  In other
words, the reduced $\chi^2$ is approximately equal to 1, indicating
that the fit of the data to the mean profile is good.  If the errors
are normally distributed, there is also a simple relationship between
$\Delta \chi^2$ above the minimum and the level of confidence that the
true period lies within the range that produce $\chi^2 \le
(\chi^2_{min}+\Delta\chi^2)$.  For models with only one free
parameter, like ours, $\Delta \chi^2 = 1$ corresponds to a 68.3\%
confidence level, and $\Delta \chi^2 = 2.71$ corresponds to a 90\%
confidence level.  If underlying red noise is present at the period of
interest, the task of identifying an oscillation and measuring its
parameters is much more difficult.  Very long baselines or repeated
observations are then needed to characterize the underlying variability.
For Z Andromedae, any red noise is at longer periods than the detected
oscillation, so we did not have this added complication.

For data with noise properties that are not precisely Gaussian, the
PDM technique can still be used, with the help of simulations to
determine the correct relationship between $\Delta \chi^2$ above the
minimum and confidence levels (van der Klis, private communication;
\cite{pre92}).  We performed such simulations for each
light curve by repeatedly injecting a fake signal with a known period
into the data, and then examining the distribution of periods
resulting from the PDM method.

The period of the oscillation as determined from each observation
individually is shown in the last column of Table 1, where the quoted
errors are roughly 68\% confidence limits.  The measurements from
observations with more than one night of observing are more precise than
single night observations because of the longer baseline.  The period
measurements from all observations are consistent, indicating that the
period was stable to within less than 15 seconds, or 1\% of the period,
for 1 year and during an outburst of the system.  More accurate
determination of the oscillation period in Z Andromedae by connecting
the data from adjacent observations will allow for important orbital
time delay measurements.  Given the system inclination of $i=47^\circ$
and taking the total system mass to be $M_{\em tot} = 2 \msun$, which is
typical for a symbiotic (\cite{sch97}, and references therein), the light
travel time across the WD orbit is $(a_{\em WD} \sin i)/c \approx 12.2\,
{\rm min}\: (\sin i / 0.73)\,(M_{\em tot} / 2 \msun)^{1/3}\, (1+ M_{\em
WD}/M_{\em RG})^{-1}$, where $a_{\em WD}$ is the distance from the WD to the
center of mass, $M_{\em WD}$ is the mass of the WD, and $M_{\em RG}$ is
the mass of the red giant.

\begin{deluxetable}{ccccccccc}
\tablewidth{0pt}
\tablecaption{Observation Log and Results\label{obslog}}
\tablehead{
\colhead{Observation} & \colhead{Date, U.T.} & \colhead{MJD} &
\colhead{Orbital} & \colhead{Obs. Length} & \colhead{$t_{exp}$} &
\colhead{$\Delta t$} & \colhead{Count Rate} & \colhead{Oscillation} \\
\colhead{Number} & \colhead{} & \colhead{$- 2450000$} &
\colhead{Phase\tablenotemark{a}} & \colhead{(hrs)} &
\colhead{(sec)} & \colhead{(sec)} & \colhead{($\times 10^4$c/s)} & \colhead{Period (sec)} }
\tablecolumns{9}
\scriptsize
\startdata
1 &1997 Jul 8 & 637 & 0.50 & 3.7 & 18 & 40 & 2.9 & $1676 \pm 19$\nl
2 & 1997 Aug 2 & 663 & 0.54 & 6.0 & 30 & 58 & 2.3 & $1686 \pm 11$\nl
3 & 1997 Aug 30 & 691 & 0.58 &  6.5 & 30 & 57 & 1.9 & $1695 \pm 15$ \nl
4 & 1997 Sep 13 & 705 & 0.59 & 8.0 & 38 & 63 & 1.9 & $1682.2 \pm 0.7$ \nl
& 1997 Sep 14 & 706 & 0.60 & 5.1 & 65 & 90 & &  \nl
5 & 1997 Oct 5 & 727 & 0.62 & 6.5 & 50 & 78 & 1.6 & $1684.2 \pm 1.5$ \nl
 & 1997 Oct 6 & 728 & 0.62 & 6.2 & 52 & 79 & & \nl
6 & 1997 Nov 1 & 754 & 0.66 & 6.2 & 40 & 68 & 1.4 & $1682.7 \pm 1.0$\nl
 & 1997 Nov 2 & 755 & 0.66 & 7.7 & 65 & 93 & & \nl
 & 1997 Nov 3 & 756 & 0.66 & 7.1 & 40 & 68 & & \nl
7 & 1997 Dec 2 & 785 & 0.70 & 5.8 & 110 & 138 & 1.2 & $1679 \pm 16$ \nl
8 & 1998 Jun 28 & 993 & 0.97 & 3.4 & 200 & 228 & 0.5 & $1682.0 \pm 3$ \nl
 & 1998 Jun 29 & 994 & 0.98 & 3.7 & 200 & 228 \nl
\enddata
\tablenotetext{a}{From the Formiggini \& Leibowitz (1994) ephemeris:
$Min(vis)= JD 2442666(\pm10) + 758.8(\pm2)E$, where $E$ is the number of
orbital cycles.} 
\tablecomments{$t_{exp}$ is the integration time, and $\Delta t$ is the
time between integration starts, which is equal to the integration time plus
the CCD readout and processing time.}
\end{deluxetable}

The peaks at twice the fundamental frequency in the early data indicate
that the pulse profile deviates from a sinusoid.  Pulse profiles created
by folding the light curves from observations \#3, \#4, and \#7 at 1683
seconds are shown are Figure 4.  The pulse fraction decreased
monotonically as the outburst decayed, from $\approx 5$ mmag
peak-to-peak in 1997 July and August (observations \#1 -- \#3) to
$\approx 2$ mmag peak-to-peak in 1997 December (observation \#7).  In
1998 June the oscillation was detected at the 2 mmag level.

\epsfig{file=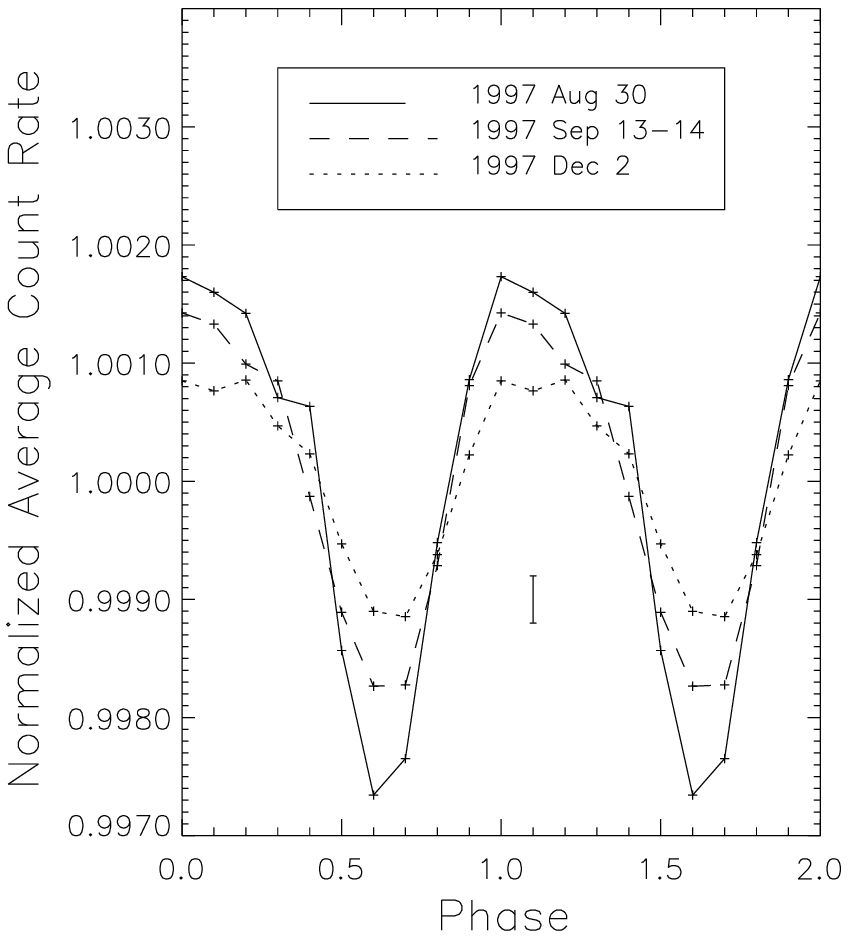}
\figcaption[f4.eps]{Z Andromedae pulse profiles at $P=1683$ s from
1997 July 7, 1997 August 30, and 1998 June 28-29.  Error bars are left off
for clarity of viewing, but 1 sigma is about 0.0003 in the normalized
units plotted, as shown by the mark in the center of the plot.  The data
are repeated for clarity.  The decrease in pulse amplitude for later
observations is significant.  \label{fig4}}

\section{The Case for Magnetic Accretion}\label{sec:magacc}

We interpret the 28-minute oscillation as the result of rotation of a
white dwarf which has a strong enough magnetic field to channel the
accretion flow onto its magnetic polar caps, as in the DQ Herculis
systems (\cite{pat94}).  Non-radial g-mode pulsations of a hot WD that
is similar to a planetary nebula nucleus (PNN) is another plausible
explanation of the oscillation, especially since g-mode pulsations with
periods close to 28 minutes have been observed in several PNN (Ciardullo
\& Bond 1996). However, these systems are multiperiodic and have
frequencies that change on month long time scales. The Z Andromedae
emission oscillated at only a single, constant frequency for an entire
year, as well as throughout an outburst during which conditions in the
WD envelope presumably changed significantly. Therefore, we conclude
that WD g-mode pulsations are unlikely to be the cause of the
oscillation.  The period of the oscillation is too long to be due to an
acoustic (p-mode) pulsation in a WD, and too short to be due to a
g-mode pulsation in a main sequence star with $M\approx 0.65 \msun$.  A
p-mode pulsation in a main sequence star is not formally ruled out, but
again one would expect more than a single mode to be present.
Therefore, the period, its stability and coherence, and the fact that
only one period is detected all support the WD magnetic dipole rotator
model.

The minimum magnetic field strength at the dipolar cap, $B_{\rm S}$,
that is needed to funnel the accretion onto the star can be
roughly estimated by requiring that the magnetospheric radius,
$r_{\rm mag}$, be larger than the white dwarf radius, $R$.  At the
magnetospheric radius, the magnetic field pressure is comparable to 
to the ram pressure of the in-falling material, giving the standard
$r_{\rm mag}\approx (\mu^4/2GM_{\em WD}\dot M^2)^{1/7}$, where $\mu=B_{\rm S}R^3/2$
is the magnetic dipole.
This then leads to a minimum field
\begin{eqnarray}
B_{\rm S} & \gtrsim  & 3 \times 10^4 {\rm G} \left(\frac{10^9 {\rm
cm}}{R}\right)^{5/4} \left(\frac{\dot M}{10^{-8}
\msyr}\right)^{1/2} \\ 
\nonumber & & \mbox{\hspace{4cm}} \times
\left(\frac{M_{\em WD}}{0.65 \msun}\right)^{1/4}. 
\end{eqnarray}
If the WD has been spun up so that it is in a rotational equilibrium, with
its spin period, $P_{\rm s}$, equal to 
the Kepler period at $r_{{\rm mag}}$, then the magnetospheric radius would be  
\begin{eqnarray}
r_{{\rm mag}} & \approx & \left(\frac{GM_{\em WD} P_{{\rm
s}}^2}{4\pi^2}\right)^{1/3} 
\\ \nonumber & \approx &
0.26\rsun\left(\frac{M_{\em WD}}{0.65 \msun}\right)^{1/3} \left(\frac{P_{{\rm s}}}{28\ {\rm
min}}\right)^{2/3},
\end{eqnarray}
and the field strength needed to have the magnetosphere at this radius
is roughly 
\begin{eqnarray}
B_{\rm S} & \approx  & 6\times10^6 {\rm G} \left(\frac{\dot M }{10^{-8} \
\msyr}\right)^{1/2} \left(\frac{10^9 {\rm cm}}{R}\right)^3 \\ 
\nonumber & & \mbox{\hspace{4cm}}
\times \left(\frac{M_{\em WD}}{0.65 \msun}\right)^{5/6}.  
\end{eqnarray}  
The time it takes for the white dwarf to reach this rotational
equilibrium is $t_{\rm spin-up}=2\pi I /NP_{\rm s}\approx 5\times 10^4 \
{\rm yr} \ (28 \ {\rm min}/ P_{\rm s})(10^{-8} \ \msyr /\mdot) $, where
$I\approx M_{\em WD}R^2/5$ is the WD moment of inertia and $N=\dot
M(GM_{\em WD}r_{{\rm mag}})^{1/2}$ is the accretion torque.  This
spin-up time is shorter than the lifetime of the red giant, so it is
likely that the system has reached this form of equilibrium.

\section{Conclusions and Implications for the Outburst Mechanism}\label{sec:obmodels}

Our survey has so far yielded one persistent periodic
oscillator.  The oscillation was detected on all 8 occasions when the
source was observed over the course of one year, and the period,
$P=1682.6 \pm 0.6$ s, was stable to within our measurement errors.  We
interpret this oscillation in terms of magnetic accretion onto a
rotating WD.  This detection is the first of its kind for a symbiotic,
and it comes from an object, Z Andromedae, in which no other phenomena
thought to be associated with magnetism have been observed.  Outburst
mechanisms need to be reconsidered in light of this discovery, and as
we now elaborate, accretion disk instabilities look to be a promising
source for the outbursts. 

The detection of an oscillation that originates at the WD surface during
an outburst has serious consequences for models of the outburst
mechanism in Z Andromedae.  Most models (\cite{mk92}, and references
therein) invoke dramatic expansion of the WD photosphere, for example as
the result of a thermonuclear shell flash or a change in $\dot M$ onto a
nearly stably burning hydrogen layer.  Evidence for such expansion and
the subsequent decrease in the effective temperature of the WD includes
a decrease in the strength of high ionization state emission lines, line
broadening, increased opacity as measured by line ratios, the appearance
of an A-F-type spectrum, and direct luminosity estimates, all during
outburst (\cite{fc95}, \cite{mk96}).  Miko{\l}ajewska \& Kenyon (1996)
deduced that the radius of the hot component increased by a factor of
$\sim100$ during previous outbursts of Z Andromedae.  However, they also
noted a few problems with the shell flash/photospheric expansion model.
The HeII emission lines evolve in a different manner than other emission
lines during outbursts.  Therefore, the outburst spectra are
inconsistent with an evolving single temperature model for the WD.
Another problem for thermonuclear runaway and steady burning shell
expansion models is the time scales.  It is difficult to reconcile
theoretical photospheric expansion time scales and shell flash
recurrence time scales with the observations (\cite{mk92}, and
references therein), especially for a low-mass WD (although see Sion \&
Ready [1992]).  Our detection of an oscillation from a region that would
be hidden by an expanded photosphere is another phenomenon that is
difficult to reconcile with models involving photospheric expansion.

Our observations do not provide information about the temperature
evolution of the hot component, so it is possible that the 1997 outburst
was significantly different from previous outbursts.  The 1997 outburst
was smaller and more asymmetric than either of two well-studied
outbursts in 1984 and 1986, which rose to $V\approx 9.6$ and $V\approx
9.1$ respectively, compared to $V\approx 9.7$ for the 1997 outburst.
Based upon multiwavelength observations, Fern\'{a}ndez-Castro
et al. (1995) suggested that the 1984 and 1986 events were
similar, but that a less massive shell was ejected during the 1984
outburst.  The 1984 event produced a smaller increase in opacity
(\cite{fc95}), so a correlation between $V$ at the outburst peak and the
nature of the outburst might exist.

   Another outburst mechanism that has been discussed for SS, although
usually not for systems that contain WDs, is thermal accretion-disk
instabilities (DI; \cite{dus86a}b) like those that lead to dwarf novae
eruptions in CVs (\cite{osa96}).  DI models have not been considered
prime candidates for explaining the outbursts in WD SS for several
reasons.  First of all, there is little direct evidence for disks around
WDs in SS.  Disks are not needed in spectral fits (\cite{mur91}) and
double peaked line profiles cannot be definitively linked to disk
emission (\cite{rob94}).  Furthermore, disk instabilities alone may not
provide sufficient energy to explain the observed flux increases
(\cite{ken86}).  Disk instability models can, however, produce time
scales that are more in accordance with the durations and recurrence
times seen in SS outbursts than thermonuclear runaway models.  We will
explore the possibility that disk instabilities play an important role
in SS outbursts more fully in a separate paper.

   There are several points that are important to note here, however.
Most importantly, if a large disk does exist around the WD in Z
Andromedae, it would be thermally unstable (\cite{dus86b};
\cite{mh92}).  Secondly, during a DI-induced outburst, the emission
region close to the WD could remain exposed, as appears to be the case
during the most recent outburst of Z Andromedae.  Finally, the
presence of a quasi-steady burning layer on the WD may affect the
energetics of the outburst resulting from a disk instability.

\acknowledgments

We would like to thank W. Ho for help with the observations, as well as
D. Chakrabarty, M. Eracleous, M. van der Klis, and G. Ushomirskiy for
useful discussions.  The work of W. Deitch modifying the timing system
at the Nickel telescope was also greatly appreciated, as was the
assistance of T. Misch and R. Stone.  This work was supported by the
California Space Institute (CS-45-97) and by a Hellman Family Faculty
Fund (UC-Berkeley) award to L. B..


\footnotesize
\begin{thebibliography}{}

\bibitem[Angel, Borra, \& Landstreet 1981]{ang81}Angel, J. R. P., Borra, E. F. \&
Landstreet, J. D. 1981, \apjs, 45, 457

\bibitem[Chanmugam 1992]{cha92}Chanmugam, G. 1992, \araa, 30, 143 

\bibitem[Ciardullo \& Bond 1996]{cia96}Ciardullo, R., \& Bond,
H. E. 1996, \aj, 111, 2332.   

\bibitem[Dobrzycka et al. 1996]{dob96}Dobrzycka, D., Kenyon, S. J., \&
Milone, A. E. 1996, \aj, 111, 414 

\bibitem[Duschl 1986a]{dus86a}Duschl, W. J. 1986a, \aap, 163, 56

\bibitem[Duschl 1986b]{dus86b}Duschl, W. J. 1986b, \aap, 163, 61

\bibitem[Fern\'{a}ndez-Castro et al. 1988]{fc88}Fern\'{a}ndez-Castro,
T., Cassatella, A., Gimen\'{e}z, A., \& Viotti, R. 1988, \apj, 324, 1016

\bibitem[Fern\'{a}ndez-Castro et al. 1995]{fc95}Fern\'{a}ndez-Castro,
T., Gonz\'{a}lez-riestra, R., Cassatella, A., Taylor, A. R., \&
Seaquist, E. R. 1995, \apj, 442, 366

\bibitem[Formiggini \& Leibowitz 1994]{fl94}Formiggini, L., \&
Leibowitz, E. 1994, \aap, 292, 534 

\bibitem[Kenyon 1986]{ken86}Kenyon, S. J. 1986, The Symbiotic Stars (Cambridge
University Press: Cambridge) 

%\bibitem[Leedj\"{a}rv \& Miko{\l}ajewski 1995]{lee95}Leedj\"{a}rv, L., \&
%Miko{\l}ajewski, M. 1995, \aap, 300,189 

\bibitem[Liebert 1988]{lie88}Liebert, J. 1988, \pasp, 100, 1302 

\bibitem[Mattei 1998]{mat98}Mattei, J. A. 1998, Observations from the
AAVSO International Database, private communication 

\bibitem[Meyer-Hofmeister 1992]{mh92}Meyer-Hofmeister, E. 1992, \aap,
253, 459

\bibitem[Michalitsianos et al. 1993]{mic93}Michalitsianos,
A. G. et al. 1993, \apj, 409, L53  

\bibitem[Miko{\l}ajewska \& Kenyon 1992]{mk92}Miko{\l}ajewska, J., \&
Kenyon, S. J. 1992, \mnras, 256, 177

\bibitem[Miko{\l}ajewska \& Kenyon 1996]{mk96}Miko{\l}ajewska, J., \& Kenyon,
S. J. 1996, \aj, 112, 1659 

\bibitem[Miko{\l}ajewski et al. 1990a]{mski90a}Miko{\l}ajewski,
M., Miko{\l}ajewska, J., Tomov, T., Kulesza, B., \& Szczerba, R. 1990a,
Acta Astronomica, 40, 129

\bibitem[Miko{\l}ajewski {\em et al} 1990b]{mski90b}Miko{\l}ajewski, M.,
Miko{\l}ajewska, J., \& Khudyakova, T. N. 1990b, \aap, 235, 219

\bibitem[Miko{\l}ajewski \& Miko{\l}ajewska 1988]{mm88}Miko{\l}ajewski,
M., \& Miko{\l}ajewska, J. 1988, in ``The Symbiotic Phenomena'',
eds. Miko{\l}ajewska, J., Friedjung, M., Kenyon, S., \& Viotti,
R. (Kluwer Academic Publishers: Dordrecht), p. 233

\bibitem[Murset et al. 1991]{mur91}Murset, U., Nussbaumer, H., Schmid,
H. M., \& Vogel, M. 1991, \aap, 248, 458 

\bibitem[Osaki 1996]{osa96}Osaki, Y. 1996, \pasp, 108, 39

\bibitem[Patterson 1994]{pat94}Patterson, J. 1994, \pasp, 106, 209 

\bibitem[Press et al. 1992]{pre92}Press, W. H., Teukolsky, S. A.,
Vetterling, W. T., \& Flannery, B. P. 1992, Numerical Recipes in
Fortran, 2nd Ed. (Cambridge University Press: New York)

\bibitem[Robinson et al. 1994]{rob94}Robinson, K., Bode, M. F.,
Skopal, A., Ivison, R. J., \& Meaburn, J. 1994, \mnras, 269, 1

\bibitem[Schmid \& Schild 1997]{sch97}Schmid, H. M., \& Schild, H. 1997,
\aap, 327, 219 

\bibitem[Sion et al. 1988]{sio88}Sion, E. M., Fritz, M. L., McMullin,
J. P. \& Lallo, M. 1988, \aj, 96, 251  

\bibitem[Sion \& Ready, 1992]{sr92}Sion, E. M., \& Ready, C. J. 1992,
\pasp, 104, 87

\bibitem[Seaquist, Taylor, \& Button 1984]{stb}Seaquist, E. R., Taylor, A. R., \&
Button, S. 1984, \apj, 284,202 

\bibitem[Stellingwerf 1978]{ste78}Stellingwerf, R.F. 1978, \apj, 224,
953  

\bibitem[Tomov et al. 1992]{tom92}Tomov, T., Zamanov, R., Kolev, D.,
Georgiev, L., Antov, A., Miko{\l}ajewski, M., \& Esipov, V. 1992,
\mnras, 258, 23 

\bibitem[Yungelson et al. 1995]{yun95}Yungelson, L., Livio, M., Tutukov,
A. \& Kenyon, S. J. 1995, \apj, 447, 656 

\end{thebibliography}
\end{document}